\documentclass{PoS}

\usepackage{latexsym}
\usepackage{amsmath}
\usepackage{amssymb, graphics}
\usepackage{graphicx}
\usepackage{multirow}
\usepackage{subfigure}

\title{Systematic errors in extracting nucleon properties from lattice QCD}

\ShortTitle{Systematic errors in extracting nucleon properties from lattice QCD}

\author{Stefano Capitani, \speaker{Bastian Knippschild}\\
        Institut f\"ur Kernphysik, University of Mainz, Becher Weg 45,
        55099 Mainz, Germany\\
        E-mail: \email{knippsch@kph.uni-mainz.de}}

\author{Michele Della Morte, Hartmut Wittig \\
        Institut f\"ur Kernphysik and Helmholtz Institute Mainz,
        University of Mainz, Becher Weg 45, 55099 Mainz, Germany}

\abstract{
\vspace{-11cm}
Form factors of the nucleon have been extracted from experiment with high precision. However, lattice calculations have failed so far to reproduce the observed dependence of form factors on the momentum transfer. We have embarked on a program to thoroughly investigate systematic effects in lattice calculation of the required three-point correlation functions. Here we focus on the possible contamination from higher excited states and present a method which is designed to suppress them. Its effectiveness is tested for several baryonic matrix elements, different lattice sizes and pion masses.
\vspace{-16cm}{\begin{flushright} {\tt MKPH-T-10-29\\ HIM-2010-04 }\end{flushright}}
 }

\FullConference{The XXVIII International Symposium on Lattice Field Theory, Lattice2010\\
		June 14-19, 2010\\
		Villasimius, Italy}

\begin{document}

%
%
\section{Introduction}

The calculation of mesonic and baryonic matrix elements from lattice QCD has made good progress in the last few years. Simulations with fully dynamical quarks have reached the physical pion mass at large lattice sizes of $L>5$ fm. For many mesonic matrix elements the overall uncertainties are of a few percent and no discrepancies between experiments and theory can be detected. Baryonic matrix elements have not yet reached this accuracy. For instance the axial charge or the electric form factor are not compatible with the experimental data, even for the largest volumes and the smallest pion masses \cite{latt10alexandrou}.\\
In this study we will focus on the electro-magnetic and axial form factors of the nucleon. The matrix element of the vector current can be expressed by the Dirac- and Pauli form factors $F_1(q^2)$ and $F_2(q^2)$ in the following way:
\begin{equation}\label{vector}
\left<N(p',s') \left|  V_\mu  \right| N(p,s) \right>=\bar u(p',s')\left\{\gamma_\mu F_1(q^2)+i\frac{\sigma_{\mu\nu}q_\nu}{2m_N}F_2(q^2)  \right\}u(p,s),
\end{equation}
where $\left| N(p,s) \right>$ is the nucleon ground state with momentum $p$ and spin $s$, and $u(p,s)$ is a Dirac spinor with mass $m_N$. The momentum transfer is given by $q=p-p'$ and $\sigma_{\mu\nu}=[\gamma_\mu, \gamma_\nu]$/2. 
The matrix element of the axial current can be expressed in terms of the axial form factor $G_A(q^2)$ and the induced-pseudoscalar form factor $G_P(q^2)$:
\begin{equation}\label{axial}
\left<N(p',s') \left|  A_\mu  \right| N(p,s) \right>=\bar u(p',s')\left\{\gamma_\mu\gamma_5 G_A(q^2)+\gamma_5\frac{q_\mu}{2m_N}G_P(q^2)  \right\}u(p,s).
\end{equation}
We will neglect $G_P(q^2)$ here completely and focus on the electro-magnetic form factors and the axial form factor at zero momentum transfer, which corresponds to the axial charge $g_A$.\\
The observed discrepancy between the experimental and lattice data may be due to systematic effects. These can be lattice artifacts, finite volume effects, large pion masses and contaminations from excited states. We will focus on the excited state contributions here but plan to address the other effects later due to our lattice setup of various lattice sizes, lattices spacings and pion masses.\\
Our ensembles for computing matrix elements were generated as part of the "Coordinated Lattice Simulations" (CLS) initiative \cite{CLS}. For our measurements we use fully dynamical $\mathcal O(a)$-improved, two flavor Wilson fermions where our solver is Schwarz preconditioned and deflation accelerated \cite{Luscher:2003vf}. Here we will present data for one lattice spacing of $0.069$ fm and two lattice sizes of $64\times32^3$ and $96\times48^3$. The pion masses vary between $300$ and $900$ MeV  \cite{Capitani:2009tg,Brandt:2010ed}.
%
%
\section{The standard plateau method}
The two-point function in Euclidean space time for a nucleon is defined as \cite{Leinweber:1990dv}:
\begin{equation}\label{twopoint}
C_2(\vec p, t)= \sum_{\vec x} e^{-i\vec p \vec x} \Gamma^P_{\mu\nu}\left< J_\nu(t, \vec x)\bar J_\mu(0)\right> 
\xrightarrow{t\rightarrow\infty} \frac{Z_B^2}{2E_p}e^{-E_p t}\operatorname{Tr}\left[\Gamma^P\left(-i p\!\!\!/  + m \right)\right].
\end{equation}
We use the interpolating field $J_\gamma = \epsilon^{abc}(u^aC\gamma_5 d^b)u_\gamma^c$ to create a nucleon where $C$ represents the charge conjugation matrix, greek letters Dirac indices and latin letters color indices. The energy $E_p$ of a nucleon with momentum $p$ is related to its mass via the dispersion relation $E^2=m^2+p^2$. The factor $Z_B$ is the coupling strength of the baryon-state with the vacuum. For the polarization matrix $\Gamma^P$ we use $\frac{1}{4}(1+\gamma_0)(1-i\gamma_3\gamma_5)$ which projects the nucleon to positive parity and polarizes it in the $z$-direction. To optimize the overlap of the nucleon correlation function with the ground state Jacobi-smearing \cite{Allton:1993wc} with HYP-smeared links in the spatial Laplacian \cite{Hasenfratz:2001hp} is used.\\
The computation of the three-point functions is more involved than for the two-point functions due to the necessity of computing extended propagators. For the vector and the axial vector form factors the following two diagrams contribute: \\[-7ex]
\begin{figure}[h]
\begin{equation}\label{3pt1}
\Sigma_d(\vec y,0;t_s,\vec p' )=\sum_{\vec x}e^{i\vec x\vec p'}\otimes\!\!\!\!\!\!\!\!\!\!\!
\parbox{4cm}{\includegraphics[angle=-90,scale=0.15]{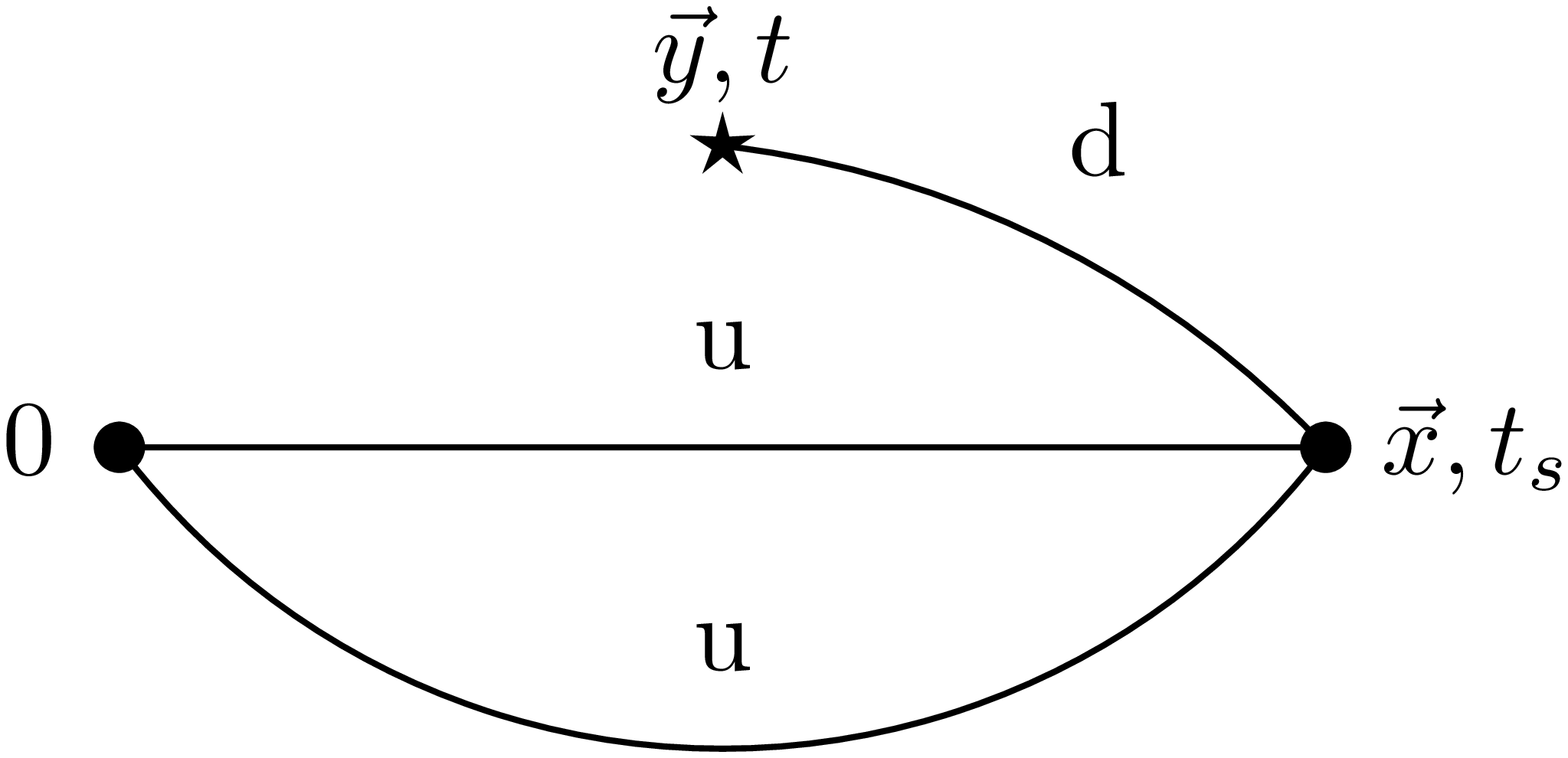}},
\end{equation}\\[-13ex]
\begin{equation}\label{3pt2}
\Sigma_u(\vec y,0;t_s,\vec p')=\sum_{\vec x}e^{i\vec x\vec p'}\otimes\!\!\!\!\!\!\!\!\!\!\!
\parbox{4cm}{\includegraphics[angle=-90,scale=0.15]{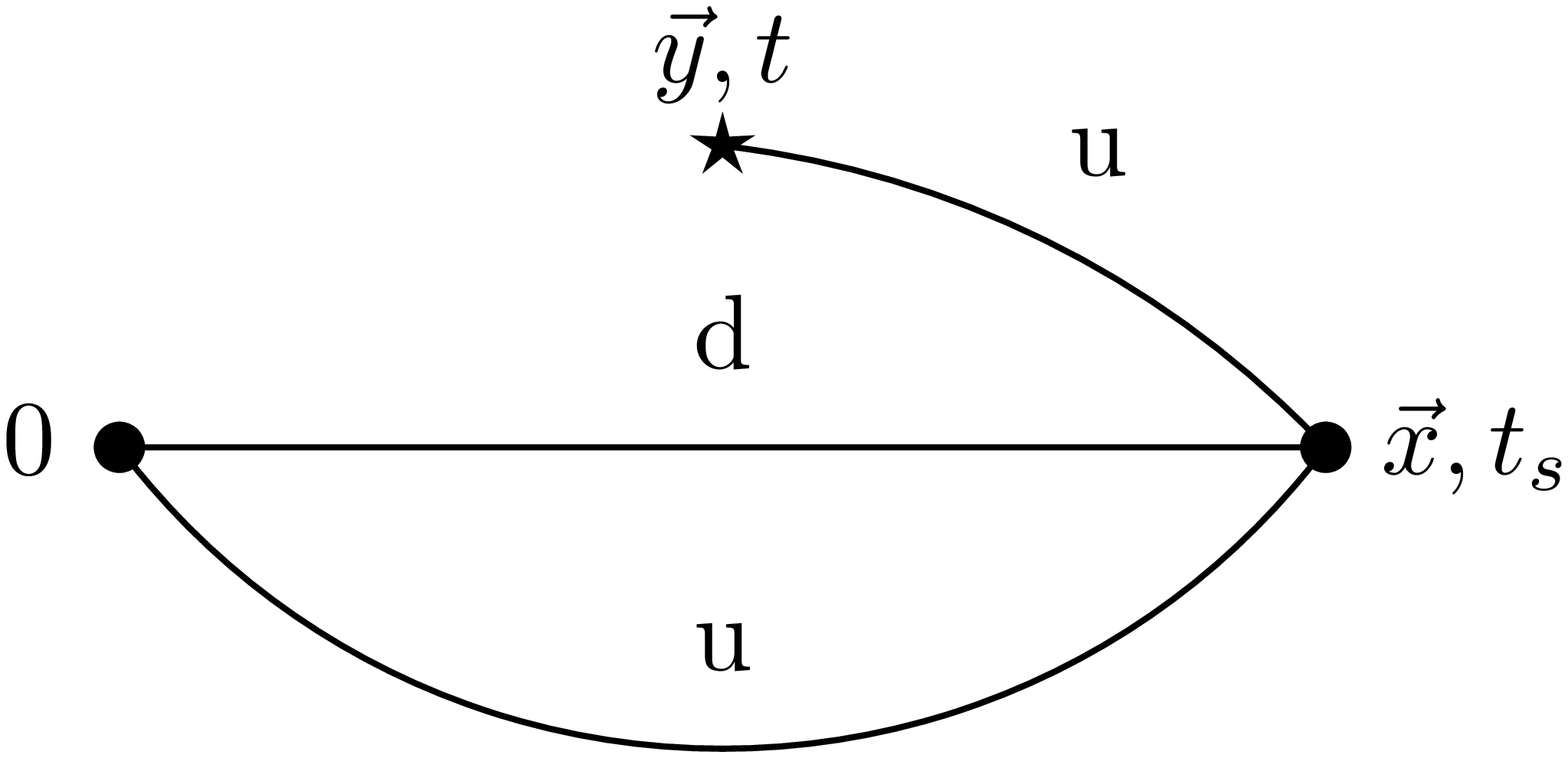}}.
\end{equation}
\end{figure}\\[-7ex]
The quark propagators are contracted at a fixed sink timeslice $t_s$ and the outgoing momentum $\vec p'$ is induced via a Fourier transformation. The initial momentum is always set to zero so the momentum transfer carried by the photon is $q=-p'$. This object is then used as a source for a new set of inversions to create a so called extended propagator \cite{Martinelli:1988rr}.\\
The three-point function can be constructed from the extended propagators as
\begin{equation}
C_3(\vec q, t, t_s) = \sum_{\vec y} Tr\left[\Gamma^{P} \left( \Sigma_u(0,\vec y)\pm\Sigma_d(0,\vec y) \right) O(y) S(y,0) \right]e^{i\vec q\vec y} \label{3ptcont1},
\end{equation}
where the plus sign corresponds to the isoscalar form factor and the minus sign to the isovector form factor. Here we focus on isovector form factors where no disconnected contributions arise. The usual quark propagator $S(y,0)$ closes the Feynman diagrams of eq.~\ref{3pt1} and \ref{3pt2} and the operator $O$ is inserted at timeslice $t$. \\
At the hadronic level and for large Euclidean time the correlation function can be written as 
\begin{equation}
C_3(\vec q, t, t_s)= \sum_{s,s'}e^{-m_Nt}e^{-E_{p'}(t_s-t)}Z_B^2\sqrt\frac{m_N}{E_{p‘}}\operatorname{Tr}\left[ \Gamma^P u(p',s')\left<N(p',s') \left|  O  \right| N(0,s) \right>\bar u(0,s) \right]\label{3ptcont2},
\end{equation}
where the matrix element $\left<N(p',s') \left|  O  \right| N(0,s) \right>$ can be expanded in terms of the form factors as shown in eq.~\ref{vector} and \ref{axial}. We use in our simulations the local currents $V_\mu(x)=\overline\Psi(x)\gamma_\mu\Psi(x)$, $A_\mu(x)=\overline\Psi(x)\gamma_5\gamma_\mu\Psi(x)$ where $\Psi(x)$ is a $u$- or $d$-quark spinor.  The local currents need to be renormalised and in the case of the electro-magnetic form factors this can be done by requiring $G_E(0)=1$ which imposes charge conservation. For the axial current we use the non pertubatively computed renormalisation constant $Z_A$ from \cite{Della Morte:2005rd}. \\
Matrix elements computed with $\mathcal{O}(a)$-improved Wilson fermions are not automatically $\mathcal{O}(a)$-improved. An improvement term has to be added to the axial current, but we are interested in the axial charge only which is extracted from a spatial component at zero momentum of the axial current and so the forward matrix element of the improvement term vanishes. Here we neglect the improvement  term for the vector current but it will be included in a later stage of our analysis.\\
The standard way to extract form factors from Euclidean correlation functions like eq.~\ref{3ptcont2} is to use ratios between two- and three-point functions. Our particular choice is \cite{Alexandrou:2008rp}:
\begin{equation}
 R(\vec q, t, t_s)=\frac{C_3(\vec q, t,t_s)}{C_2(\vec 0,t_s)} \sqrt{\frac{C_2(\vec q,t_s-t) C_2(\vec 0,t) C_2(\vec 0,t_s)}{C_2(\vec 0,t_s-t) C_2(\vec q,t) C_2(\vec q, t_s)}}.
\end{equation}
This ratio cancels the exponential factors and gives usually long plateaux with small errors. With our choice of the polarization matrix we get access to the following quantities:
\begin{eqnarray}
Z_V\cdot\operatorname{Re} \left[ R(\vec q, t, t_s)_{\gamma_0}\right]&=& \sqrt{\frac{m+E_q}{2E_q}} 
\left\{ F_1(q^2) - \frac{\vec p'^2}{2m(m+E_{\vec p'})} F_2(q^2) \right\}
\label{ratio1},  \\
Z_V\cdot\operatorname{Re} \left[ R(\vec q, t, t_s)_{\gamma_i}\right]_{i=1,2}  &=&\epsilon_{ij} p_j  \sqrt{\frac{1}{2E_q(E_q+m)}}
\left\{ F_1(q^2)+F_2(q^2) \right\}
\label{ratio2}, \\
Z_A\cdot\operatorname{Im} \left[R(\vec q, t, t_s)_{\gamma_5\gamma_3}\right] &=&  \sqrt{\frac{m+E_q}{2E_q}} \left\{G_A(q^2)-\frac{2q_3^2}{m} G_p(q^2)\right\}\xrightarrow{\vec q \rightarrow 0}g_A\label{ratio3}.
\end{eqnarray}
The masses and energies in the pre-factors of the bare ratios can be determined from the two-point function of the nucleon. In principle the form factors can also be extracted from other components of the ratio but these turned out to be too noisy within the standard approach.\\ 
The plateau method should give the ground state value of the matrix elements, but higher state contributions can lead to wrong plateau values.
\begin{figure}[h]
\centering 
\subfigure{\includegraphics[scale=0.5]{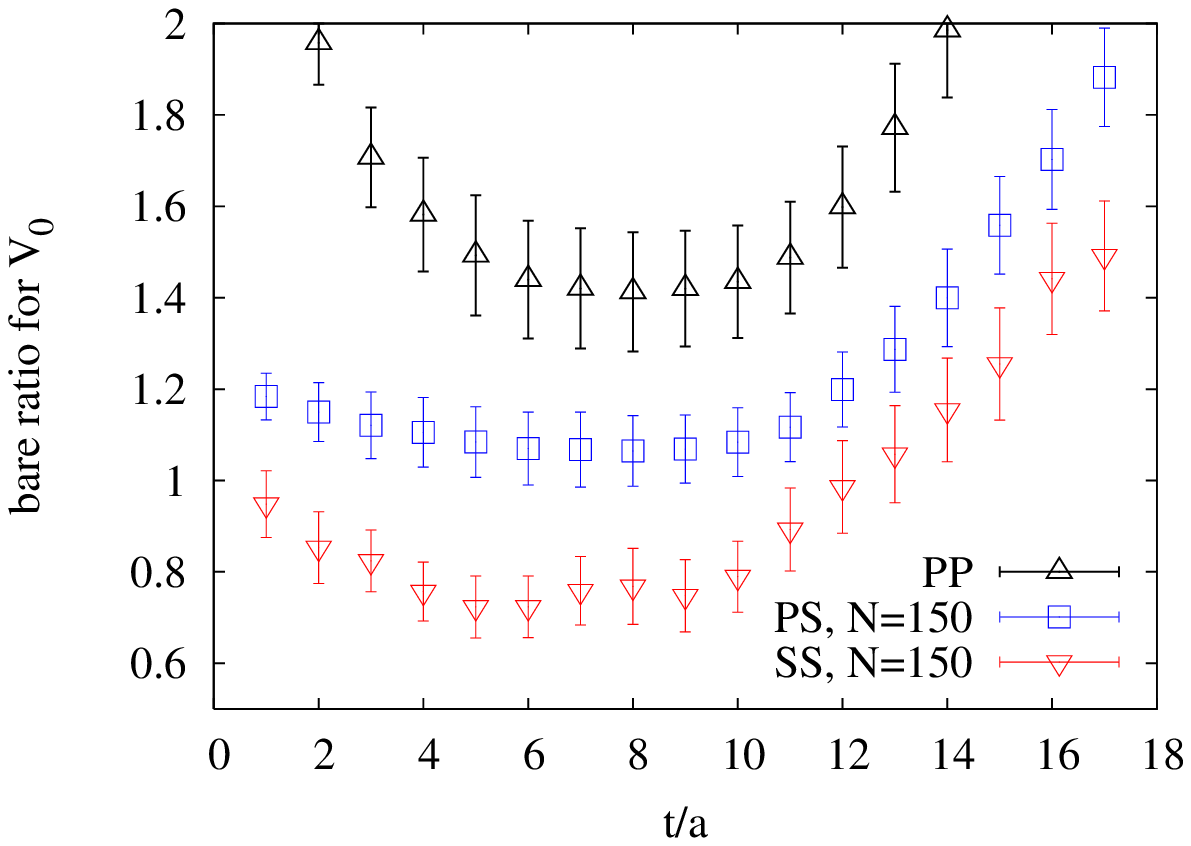}}\label{badplots}
\subfigure{\includegraphics[scale=0.5]{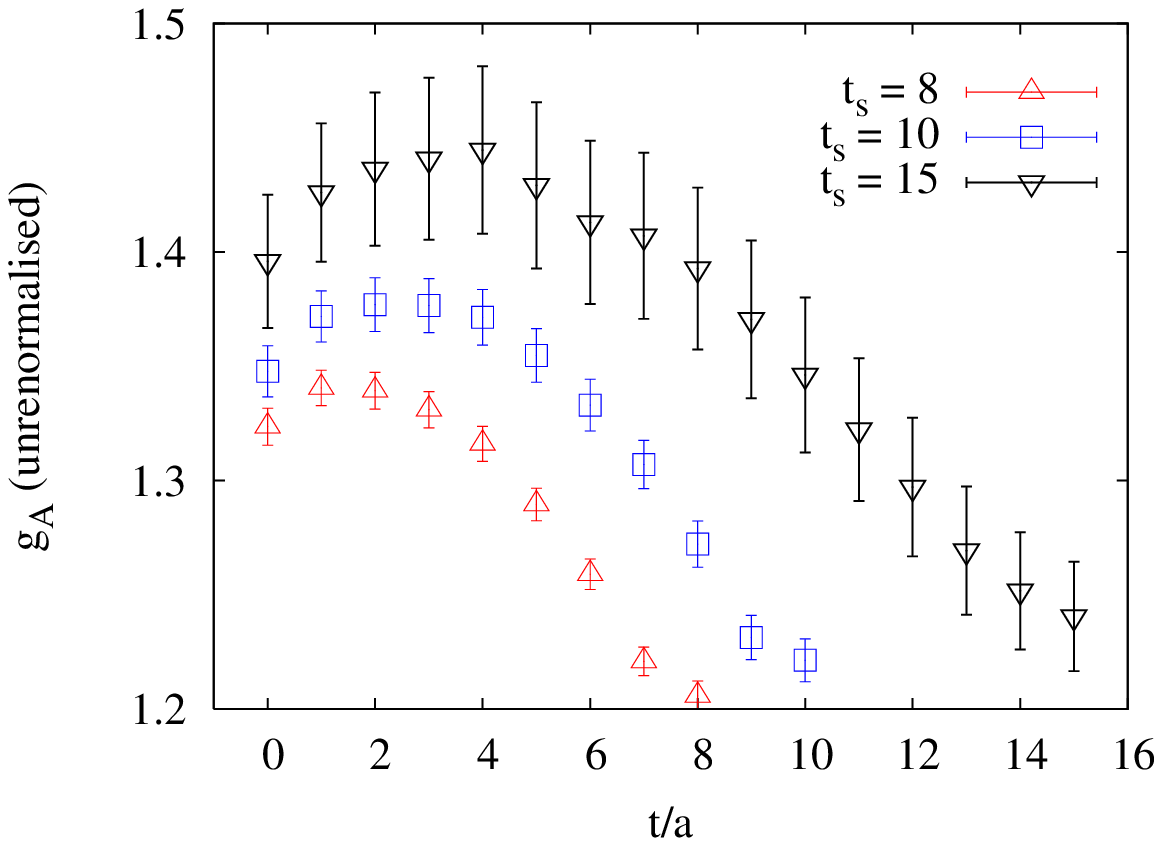}}
\\[-1ex]\caption{Examples for the standard method for $V_0$ and $g_A$}\label{examplespict}
\end{figure}
In the left panel of fig.~\ref{examplespict} we show the connected part of the zero component of the isoscalar vector form factor for three different source/sink combinations on a $64\times32^3$ lattice with a pion mass of $m_\pi=550$ MeV. The black points are from a point source, the purple ones from a Jacobi-smeared source and the red ones from a smeared source/sink combination. All three graphs give reasonable plateaux yet they are not compatible within statistical errors. \\
In the right panel the ratio for the axial charge for three different sink positions $t_s$ and a Jacobi-smeared source on a $64\times32^3$ lattice with a pion mass of $m_\pi=415$ MeV is shown. The ratio should be independent of the position of the sink but it is obviously not. Finally we cannot conclude that the real plateau value is obtained even for the largest sink timeslice $t_s=15$.

%
%
\section{The summation method}

\begin{figure}[h]
\centering
\includegraphics[scale=0.5]{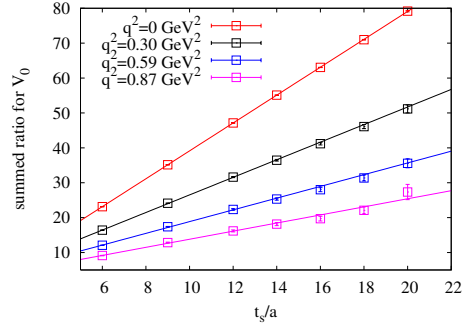}
\\[-1ex]\caption{The summation method for $V_0$ and different momenta}\label{bspsummeth}
\end{figure}

We will now argue that the large dependence of the ratios on $t_s$ and different smearing levels is mainly induced by higher order corrections and propose a method to reduce them. The ratio for an arbitrary operator can be written as:
\begin{equation}
\begin{array}{l}
R(\vec q, t, t_s)= R_G+\mathcal{O}\left(e^{-\Delta t}\right)+ \mathcal{O}\left(e^{-\Delta'(t_s-t)}\right).\label{ratiometh}
\end{array}
\end{equation}
It can be expressed in terms of the value of the ground state $R_G$ and the exponentially suppressed excited state contributions. The energy gaps $\Delta$ and $\Delta'$ can be different for different quantum numbers at source and sink. Assuming these gaps come from a two-pion state these contributions are large for small pion masses and small $t_s$.\\
To deal with these effects an alternative method can be used known as the summation method \cite{Maiani:1987by}. The basic idea is to sum the ratio in $t$ up to $t_s$. Doing so results in
\begin{equation}
\begin{array}{l}
\sum\limits_{t=0}^{t_s} R(\vec q, t, t_s) = R_G \cdot t_s+c(\Delta, \Delta')+\mathcal{O}\left(t_s e^{-\Delta t_s}\right)+ \mathcal{O}\left(t_s e^{-\Delta' t_s}\right) \label{summeth},
\end{array}
\end{equation}
 where again only the first excited state is taken into account.
The ground state can be extracted from the slope of a linear function in $t_s$. Also here an exponantially suppressed higher order correction survives but in contrast to the standard method it is only $t_s$-dependent and so much smaller.\\
Fig.~\ref{bspsummeth} illustrates how well the summation method works. We plot the zero component of the connected  part of the isoscalar vector current for four different momentum transfers.
The behavior is linear as expected and no deviation from the linearity can be detected. This means that even for the smallest $t_s$ the excited state contributions are depleted.\\
\begin{figure}[h]
\centering 
\subfigure{\includegraphics[scale=0.5]{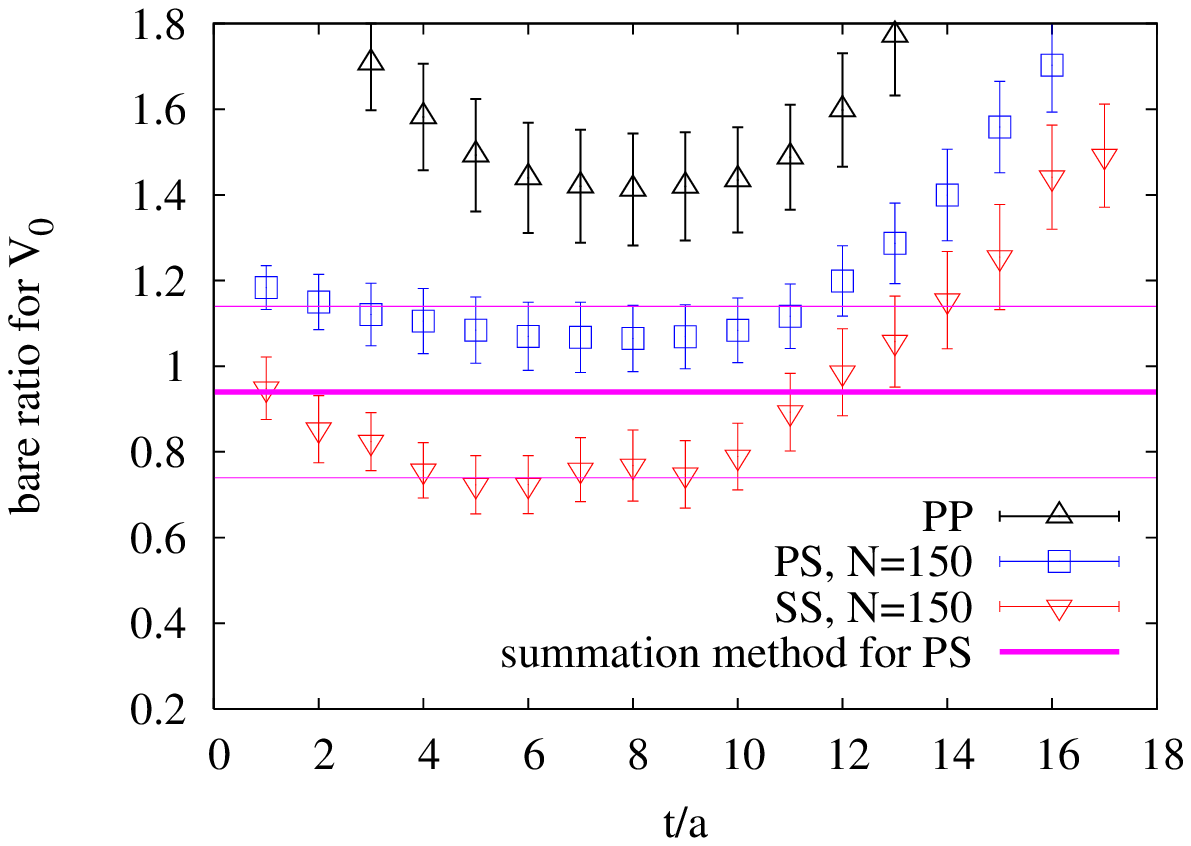}}
\subfigure{\includegraphics[scale=0.5]{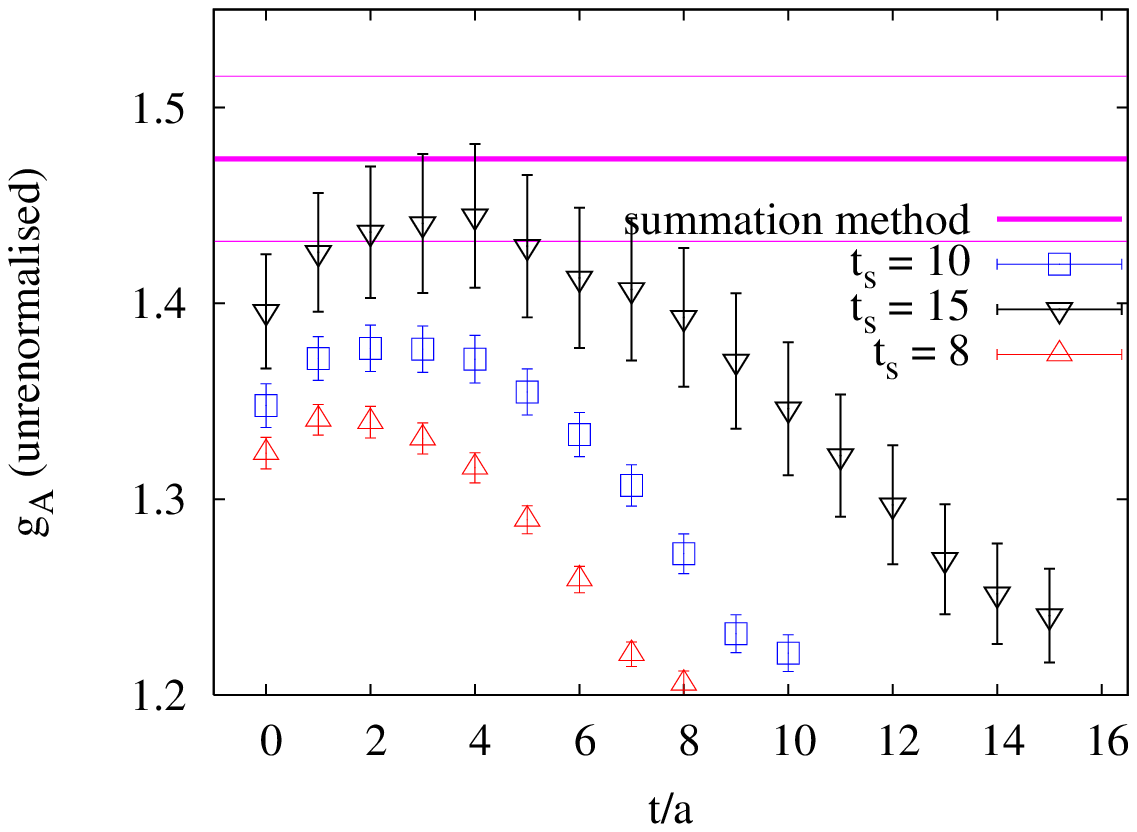}}
\\[-1ex]\caption{Examples for the standard method and the summation method for $V_0$ and $g_A$}\label{examplespics2}
\end{figure}In order to compare the summation method directly with the standard method we show in fig.~\ref{examplespics2} the graphs of fig.~\ref{examplespict} again, but now including the values from the summation method. In the left panel the purple line represents the summation method using a correlation function with a smeared source. The thick line in the middle corresponds to the mean value and the thinner outer lines show the statistical error. The summation method depletes the higher state contribution in both cases at the expense of a larger statistical error. In the right panel the ambiguity in $t_s$ is completely resolved by using the summation method (purple lines) and the trend seems to end there.\\[-3ex]
%
%
\section{Results}
Our results for the axial charge and the vector form factors are preliminary. As mentioned before the improvement term for the vector current is not implemented yet and the scale is not set to its final value so the momenta and pion masses may change. All values were extracted with the summation method.
\begin{figure}[h]
\centering \label{formfactors}
\subfigure{\includegraphics[scale=0.5]{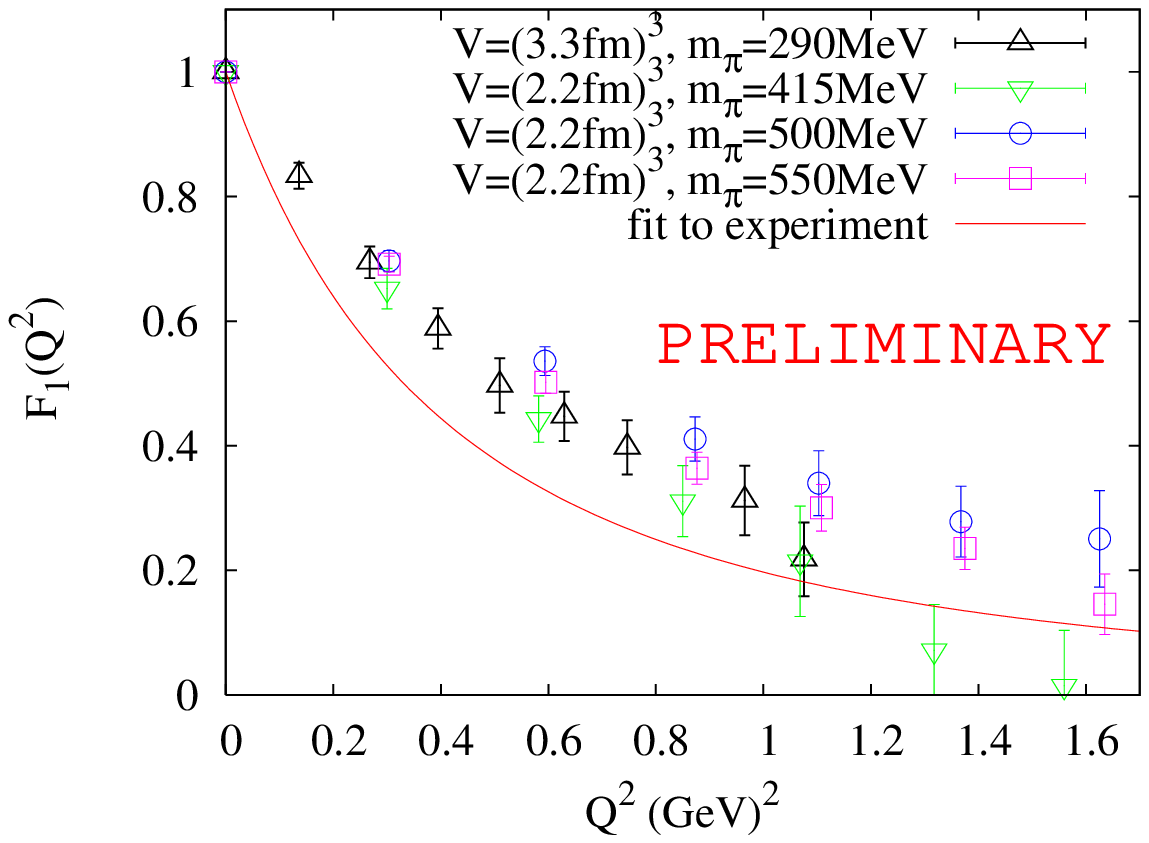}}
\subfigure{\includegraphics[scale=0.5]{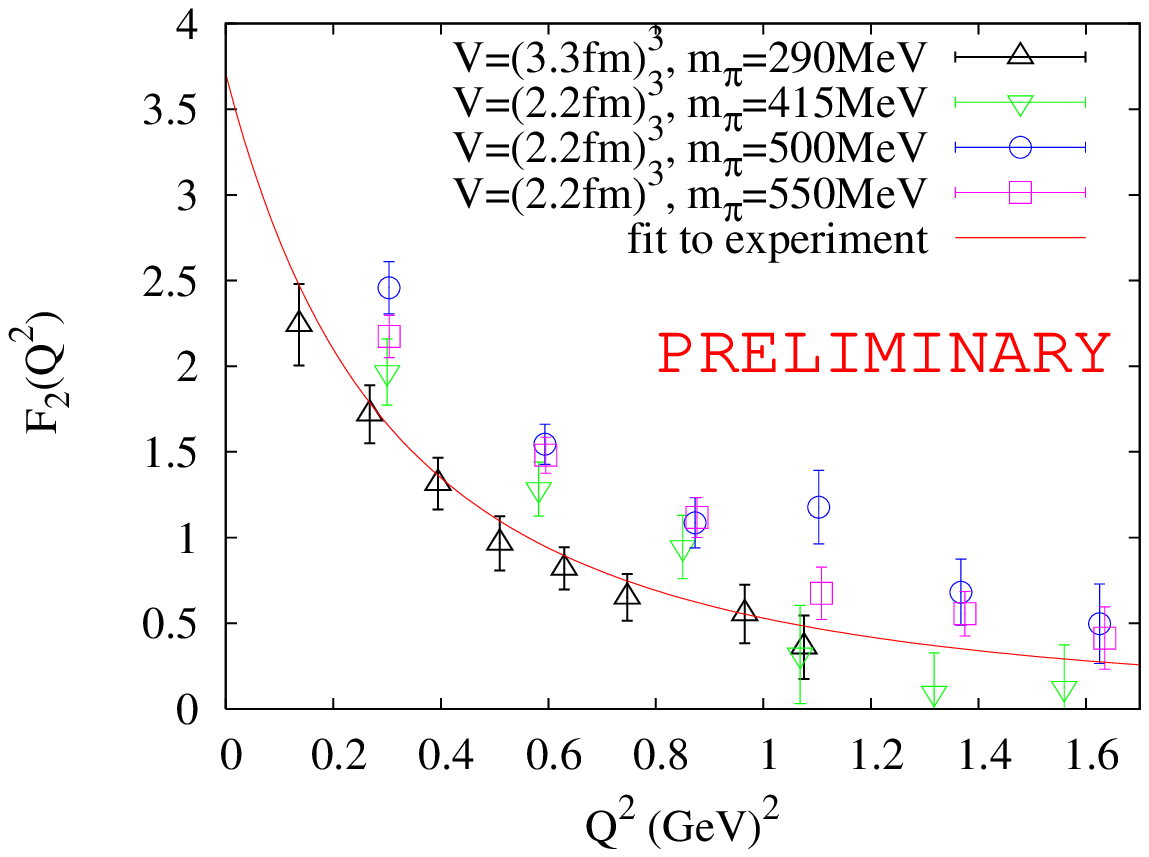}}
\\[-1ex]\caption{Preliminary results for the Dirac and the Pauli form factors extracted with the summation method.}
\end{figure}
In fig.~\ref{formfactors} we show our results for the Dirac and the Pauli form factors. While our values for the Pauli form factor for the largest lattice with the smallest pion mass is in good agreement with the experimental curve the Dirac form factor shows a different slope. So the excited state contributions seem not to be the only effect which causes the discrepancy between lattice results and experiment.\\
\begin{figure}[h]
\centering
\includegraphics[scale=0.5]{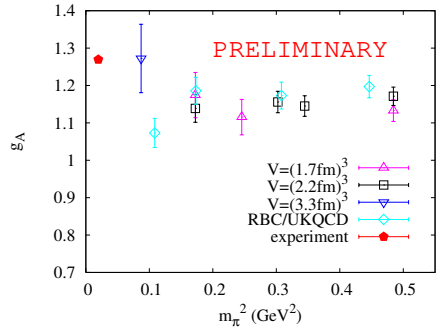}
\\[-1ex]\caption{Preliminary results for the axial charge extracted with the summation method.}\label{gA}
\end{figure}In fig.~\ref{gA} we show our data for the axial charge for different lattice sizes and pion masses. Data from the RBC/UKQCD collaboration \cite{Yamazaki:2009zq} are shown for comparison. In contrast to other collaborations we do not see a downward trend for small pion masses. The impact of the summation method is larger for small pion masses because the energy gaps $\Delta$ and $\Delta'$ decrease and the influence of excited state contributions increases.

%
%
\section{Summary and outlook}
We showed that the control of excited states is crucial and although plateaux of baryonic matrix elements extracted with the standard method may look reasonable this method is insufficient to ensure the absence of excited state contributions. Smearing appears not to be enough to suppress these contributions so we introduced an alternative method, the summation method. It is promising to help with excited state contributions but more inversions are needed and the statistical error grows compared to the standard method. To minimize these drawbacks this method will be optimized and tuned. \\
To be able to extrapolate the matrix elements to the physical point simulations with smaller pion masses and different lattice spacings $(\beta=5.2,5.5)$ are being analysed. In the analysis of the vector form factors the $\mathcal{O}(a)$ improvement terms will be included, and finally we are also interested in the full axial form factor $G_A(q^2)$ and  the induced pseudo scalar form factor $G_P(q^2)$.\\\\
\textbf{Acknowledgments:} We thank our colleagues within the CLS project for sharing gauge ensembles. 
Calculations of correlation functions were performed on the dedicated QCD platform “Wilson” at 
the Institute for Nuclear Physics, University of Mainz. This work is supported by DFG (SFB443) and
GSI. We are grateful to Harvey B. Meyer for fruitful discussions and helpful comments.

%
%


\begin{thebibliography}{99}

\bibitem{latt10alexandrou}
  C.~Alexandrou, Plenary talk at The XXVIII International Symposium on Lattice Field Theory, Lattice 2010, Villasimius, Sardinia, Italy, June, 2010
  
\bibitem{CLS}
  https://twiki.cern.ch/twiki/bin/view/CLS/WebHome

\bibitem{Luscher:2003vf}
  M.~Luscher,
  JHEP {\bf 0305}, 052 (2003)
  [arXiv:hep-lat/0310048];  \\
  JHEP {\bf 0707}, 081 (2007)
  [arXiv:0706.2298].

\bibitem{Capitani:2009tg}
  S.~Capitani, M.~Della Morte, E.~Endress, A.~Juttner, B.~Knippschild, H.~Wittig and M.~Zambrana,
  PoS {\bf LAT2009}, 095 (2009)
  [arXiv:0910.5578].

\bibitem{Brandt:2010ed}
  B.~B.~Brandt {\it et al.},
  arXiv:1010.2390.

\bibitem{Leinweber:1990dv}
  D.~B.~Leinweber, R.~M.~Woloshyn and T.~Draper,
  Phys.\ Rev.\  D {\bf 43}, 1659 (1991).

\bibitem{Allton:1993wc}
  C.~R.~Allton {\it et al.}  [UKQCD Collaboration],
  Phys.\ Rev.\  D {\bf 47}, 5128 (1993)
  [arXiv:hep-lat/9303009].

\bibitem{Hasenfratz:2001hp}
  A.~Hasenfratz and F.~Knechtli,
  Phys.\ Rev.\  D {\bf 64}, 034504 (2001)
  [arXiv:hep-lat/0103029].
  
\bibitem{Martinelli:1988rr}
  G.~Martinelli and C.~T.~Sachrajda,
  Nucl.\ Phys.\  B {\bf 316}, 355 (1989).

\bibitem{Della Morte:2005rd}
  M.~Della Morte, R.~Hoffmann, F.~Knechtli, R.~Sommer and U.~Wolff,
  JHEP {\bf 0507}, 007 (2005)
  [arXiv:hep-lat/0505026].

\bibitem{Alexandrou:2008rp}
  C.~Alexandrou {\it et al.}  [
  ETM Collaboration],
  PoS {\bf LATTICE2008}, 139 (2008)
  [arXiv:0811.0724].

\bibitem{Maiani:1987by}
  L.~Maiani, G.~Martinelli, M.~L.~Paciello and B.~Taglienti,
  Nucl.\ Phys.\  B {\bf 293}, 420 (1987).

\bibitem{Yamazaki:2009zq}
  T.~Yamazaki {\it et al.},
  Phys.\ Rev.\  D {\bf 79}, 114505 (2009)
  [arXiv:0904.2039].



\end{thebibliography}
\end{document}